\documentclass[preprint,showpacs,preprintnumbers,amsmath,amssymb,eqsecnum]{revtex4}

\usepackage{graphicx}
\usepackage{dcolumn}
\usepackage{bm}

\def\Schrodinger{Schr\"odinger }
\def\B{{\cal B}}
\def\Q{{\cal Q}}
\def\S{{\cal S}}
\def\E{{\cal E}}
\def\F{{\cal F}}

\def\<{\langle}
\def\>{\rangle}
\def\s1{{\bf S}^1}
\def\A{{\bf A}}

\def\C{{\cal C}}
\def\Ham{{\bf H}}
\def\vac{{\rm vac}}
\def\X{{\bf X}}

\begin{document}

\title{Functional Time Evolution, Anomaly Potentials,\\ and the Geometric Phase}

\author{C.~G.~Torre}

 \affiliation{Department of Physics, Utah State University, Logan, Utah 84322-4415}

\begin{abstract}

A free quantum field in 1+1 dimensions admits unitary \Schrodinger picture dynamics along any foliation of spacetime by Cauchy curves.   Kucha\v r showed that the \Schrodinger picture state vectors, viewed as functionals of spacelike embeddings, satisfy a functional \Schrodinger equation in which the generators of time evolution are the scalar field energy-momentum densities with a particular normal-ordering and with a (non-unique) $c$-number contribution. The $c$-number contribution to the \Schrodinger equation, called the ``anomaly potential'', is needed to make the equation integrable in light of the Schwinger terms present in the commutators of the normal-ordered energy-momentum densities.   Here we give a quantum geometric interpretation of the anomaly potential. In particular, we show the anomaly potential corresponds to the expression in a gauge of the natural connection on the bundle of vacuum states over the space of embeddings of Cauchy curves into the spacetime.  The holonomy of this connection is the geometric phase associated with dynamical evolution along a closed path in the space of embeddings generated by the normal-ordered energy-momentum densities.  The presence of the anomaly potential in the functional \Schrodinger equation provides a dynamical phase which removes this holonomy, so that there is no net phase change for quantum transport around closed loops in the space of embeddings.

\end{abstract}


\maketitle

\section{\label{sec:section1}Introduction}\protect

Time evolution in a field theory is usually specified in terms of the behavior of various dynamical variables as one moves along a given foliation of spacetime by Cauchy surfaces.  One is often interested -- particularly in applications to gravitational physics -- in the possibility of considering dynamical evolution along {\it any} foliation by Cauchy surfaces. In order to consider different foliations it is convenient to view them as different curves in the space of embeddings $\cal E$ of Cauchy surfaces into spacetime.  In this setting, dynamical variables can be viewed as functionals on $\cal E$ and this more general type of dynamical evolution is often called ``functional time evolution'' \cite{Kuchar1992}.  In classical field theory functional time evolution is quite well-understood (see {\it e.g.}, Ref.~\onlinecite{Kuchar1976} and references therein). In quantum field theory considerably less is known about functional time evolution; see 
Ref.~\onlinecite{Dirac1964,Kuchar1992,TV1998,TV1999}.

For a free quantum field in a flat 1+1 dimensional spacetime it is a non-trivial fact that one can use a simple Fock representation to implement functional time evolution in the \Schrodinger picture \cite{TV1998}. Put differently, one has unitary \Schrodinger picture dynamics along any foliation of spacetime by Cauchy curves, all within a single Fock representation. In this setting the \Schrodinger picture state vector of the quantum field can be viewed as a mapping from $\cal E$ into the Fock space.  As shown in Ref.~\onlinecite{Kuchar1989, TV1998}, the state vector then satisfies a functional \Schrodinger equation in which the role of the Hamiltonian density is played by the quantized energy-momentum densities of the field. The precise definition of these densities is, however, somewhat intricate.  The energy-momentum densities are defined as the sum of (i) the normal-ordered classical expression, using an embedding-dependent set of creation and annihilation operators, and (ii) a non-unique embedding-dependent multiple of the identity operator. The need for contribution (ii) was first recognized by Kucha\v r, who called this $c$-number the ``anomaly potential'' \cite{Kuchar1989}.  This terminology arose because the $c$-number contribution is needed to make the functional \Schrodinger equation integrable in light of anomalous terms (``Schwinger terms'') present in the commutators of the energy-momentum densities. The presence of the anomaly potential can be viewed as an embedding-dependent adjustment to the naive normal-ordering definition of the \Schrodinger picture energy-momentum densities. This adjustment guarantees that dynamical evolution of the state vector between two time slices is independent of the choice of interpolating foliation.

In this note we expose the quantum theoretic origins of the anomaly potential in the \Schrodinger equation. A hint is given in Ref.~\onlinecite{Kuchar1989} where an analogy is drawn between the anomaly potential and the vector potential for a particle moving in an electromagnetic field. A geometrical interpretation of the latter situation is, of course, that the vector potential corresponds to a connection on a principal bundle over spacetime. Is there a corresponding geometrical interpretation of the anomaly potential? 
Here we give such an  interpretation for the anomaly potential  using techniques and results from the extensive investigations into the quantum geometric phase \cite{BMNKZ}.

  We consider the Fock representation used in Ref.~\onlinecite{Kuchar1989, TV1998} for a free, massless scalar field on a flat spacetime ${\bf R}\times {\bf S}^1$  (Section \ref{sec:section2}). 
The  functional \Schrodinger picture is defined by a family of unitary transformations on the Fock space parametrized by the time slice, {\it i.e.}, the embedding (Section \ref{sec:section4}). 
The Hilbert space of Fock state vectors is a principal $U(1)$ bundle $\Q \to \S$ over the manifold of Fock states; $\cal Q$ is equipped with a natural connection (Section \ref{sec:section3}).    The \Schrodinger picture images of the Heisenberg picture vacuum state define a mapping from $\E$ to $\S$.  This mapping determines an induced principal $U(1)$ bundle $\B\to \E$ (Section \ref{sec:section5}).  There is a connection on $\B$ induced from the connection on $\Q$. Its holonomy corresponds to the geometric phase associated with the transport of state vectors along closed curves in $\E$. The expression of this connection in a suitable gauge is precisely  the Kucha\v r anomaly potential (Section \ref{sec:section6}). The effect of the anomaly potential in the \Schrodinger equation is then to provide a dynamical phase which cancels the geometric phase associated with time evolution along closed paths in $\cal E$, the latter phase being generated by the normal-ordered energy-momentum densities (Section \ref{sec:section7}). From this it follows that dynamical evolution of the state vector in the \Schrodinger picture is independent of the choice of foliation connecting given initial and final time slices.

\section{\label{sec:section2} The quantum scalar field on ${\bf R}\times {\bf S}^1$}

We will consider a free, massless, scalar quantum field on a flat cylindrical spacetime $(M,\eta)$. The spacetime manifold is $M={\bf R}\times {\bf S}^1$ with coordinates $t\in (-\infty,\infty)$ and $x\in(0,2\pi)$. The spacetime metric in these coordinates is
\begin{equation}
\label{metric}
\eta = -dt\otimes dt + dx\otimes dx.
\end{equation}
We will also use null coordinates $x^\pm = t\pm x$, where
\begin{equation}
\label{ }
\eta = -\frac{1}{2}\left( dx^+\otimes dx^- + dx^-\otimes dx^+\right).
\end{equation}
The (real symplectic vector) space of smooth solutions to the wave equation is given by
\begin{equation}
 \varphi_{,+-} = 0    ,
\label{eq:KG}
\end{equation}
\begin{equation}
\varphi ={1\over\sqrt{2\pi}}\Bigg\{q + p t
+{1\over\sqrt{2}}\sum_{n=-\infty\atop n\neq0}^\infty
{1\over\sqrt{|n|}}\Big[a^{(+)}_n e^{-inx^+}
 + a^{(-)}_n e^{-inx^-}\Big]\Bigg\},
\end{equation}
where $q$ and $p$ are real and
\begin{equation}
a^{(\pm)*}_n = a^{(\pm)}_{-n},
\end{equation}
define rapidly decreasing sequences of complex numbers. Thus the classical field $\varphi$ is a superposition of a topological zero-frequency mode with left and right moving modes with frequencies $|n|=1,2,\dots$.

Following Ref.~\onlinecite{Kuchar1989, TV1998}, we define the operator representative of the field, $\hat \varphi$, in the Heisenberg picture as follows. Introduce symmetric Fock spaces $\F^{(\pm)}$ for left and right moving modes with associated annihilation and creation operators $\hat a_n^{(\pm)}$, $\hat a_n^{(\pm)\dagger}$, $n=1,2,\dots$, with non-vanishing commutators
\begin{equation}
\label{ }
[\hat a_n^{(+)}, \hat a_m^{(+)\dagger}] = \delta_{nm}\hat I^{(+)},\quad   [\hat a_n^{(-)}, \hat a_m^{(-)\dagger}] =  \delta_{nm}\hat I^{(-)}, \quad n,m>0
\end{equation}
in the usual way ($\hat I^{(\pm)}$ are the identity operators on $\F^{(\pm)}$).  Introduce a Hilbert space $L^2({\bf R})$ with elements being complex-valued functions $\psi=\psi(p)$ upon which we define
\begin{equation}
\label{ }
\hat q \psi(p) =  i{d\psi\over dp},\quad \hat p\psi =  p\psi(p).
\end{equation}
We define the Hilbert space for the field as 
\begin{equation}
\label{ }
\F = L^2({\bf R})\otimes\F^{(+)}\otimes \F^{(-)}.
\end{equation}
On $\F$ the quantum scalar field is defined as an operator-valued spacetime distribution via
\begin{equation}
\label{ }
\hat \varphi ={1\over\sqrt{2\pi}}\Bigg\{\hat q + \hat p t
+{1\over\sqrt{2}}\sum_{n=-\infty\atop n\neq0}^\infty
{1\over\sqrt{|n|}}\Big[\hat a^{(+)}_n e^{-inx^+}
 + \hat a^{(-)}_n e^{-inx^-}\Big]\Bigg\},
\end{equation}
where 
\begin{equation}
\label{ }
  \hat a_n^{(\pm)\dagger}= \hat a_{-n}^{(\pm)} .
\end{equation}
Throughout this paper we use a hat (\ $\hat{}$\ ) to denote an operator on $\F$.

For each $\psi\in L^2({\bf R})$ there is a state vector, denoted by $|\psi,0\>$, which we call ``the'' Heisenberg vacuum state, since it satisfies
\begin{equation}
\label{ }
a_n^{(\pm)}|\psi,0\> = 0,\quad n>0. 
\end{equation}
A basis for $\F$ is obtained from a basis for $L^2({\bf R})$ and from repeated applications of the creation operators to $|\psi,0\>$. 

Throughout this article all kets and bras will denote elements of $\F$ and $\F^*$.

\section{\label{sec:section4} Embeddings and functional time evolution}

We denote by $X\colon \s1\to M$ a smooth, orientation-preserving, spacelike embedding of a circle into the two-dimensional spacetime.  
%
Using coordinates $x^\alpha$ and $\sigma\in (0,2\pi)$ on $M$ and $\s1$ respectively, the embedding is defined by 2 smooth functions $X^\alpha$:
\begin{equation}
\label{ }
x^\alpha = X^\alpha(\sigma)
\end{equation}
satisfying
\begin{equation}
\label{gamma}
\gamma = \eta_{\alpha\beta}X^{\alpha\prime}\,X^{\beta\prime} >0
\end{equation}
and 
\begin{equation}
\label{ }
(dx)_\alpha X^{\alpha\prime} >0,
\end{equation}
where $dx$ is the 1-form appearing in (\ref{metric}).
Using null coordinates $x^\pm$ these two conditions are equivalent to
\begin{equation}
\label{ }
X^{+\prime}>0,\quad  X^{-\prime} <0. 
\end{equation}

For future use we note that the set of all orientation-preserving spacelike embeddings, denoted by $\E$, is an infinite-dimensional manifold  which is contractible. To see this latter feature we consider $X_0,X_1\in \E$. For any $\lambda\in[0,1]$ the quantity
\begin{equation}
\label{ }
X(\lambda) = \lambda X_1 + (1-\lambda) X_0
\end{equation}
is an orientation-preserving spacelike embedding as can easily be seen, {\it e.g.}, by working in null coordinates.  It follows that $\E$ is contractible to a point.

Although the field operator $\hat \varphi$ is defined as an operator-valued distribution on $M$, the simplicity of two dimensions allows one to restrict it and its derivatives to any embedded circle thus defining operator-valued distributions on $\s1$. In particular,
we define the canonical field operators associated to an embedding $X$ as
\begin{equation}
\label{ }
\hat \phi[X] = X^*\varphi,\quad\hat \pi[X] = \sqrt{\gamma} X^* L_n \hat \varphi,
\end{equation}
where $\gamma$ is the induced metric (\ref{gamma}) on the embedded circle and $n$ is the future-pointing unit normal to the embedded circle. In null coordinates $x^\pm$ on $M$ and using $\sigma$ as a coordinate on $\s1$ these definitions are
\begin{eqnarray}
\label{ }
\hat \phi[X](\sigma) &=& \hat \varphi(X^+(\sigma), X^-(\sigma)),\\
\hat \pi[X](\sigma) &=& X^{+\prime}(\sigma) {\partial\hat \varphi(X^+(\sigma), X^-(\sigma))\over\partial x^+} - 
X^{-\prime}(\sigma) {\partial \hat \varphi(X^+(\sigma), X^-(\sigma))\over\partial x^-}.
\end{eqnarray}
Such operators are functionals of the embeddings and can be used, in conjunction with a choice of state vector, to compute outcomes of measurements of observables associated to any specific time slice in the Heisenberg picture. 

The canonical field operators associated with any two spacelike embeddings $X_1$ and $X_2$ are unitarily related \cite{TV1998}, that is, for each pair of embeddings there exists a unitary transformation $\hat U[X_1,X_2]\colon \F\to \F$ such that
\begin{equation}
\label{eq:HU}
\hat \phi[X_2] = \hat U^\dagger[X_1,X_2]\,\hat \phi[X_1]\, \hat U[X_1,X_2],\quad \pi[X_2] = \hat U^\dagger[X_1,X_2]\, \hat \pi[X_1]\, \hat U[X_1,X_2].
\end{equation}
Thus functional time evolution is unitarily implemented in the Heisenberg picture. 

Because the Fock representation constructed in \ref{sec:section1}I is irreducible, $\hat U$ is unique up to a (possibly embedding-dependent) phase. See Ref.~\onlinecite{BMNKZ} for an explicit expression of $\hat U$.

Functional Heisenberg equations can be obtained from (\ref{eq:HU}) by viewing $X_2$ as an infinitesimal deformation of $X_1$.  We then get \cite{Kuchar1989, TV1998},
\begin{eqnarray}
\label{eq:Heq1}
{\delta \hat \phi[X](\sigma)\over\delta X^\alpha(\sigma^\prime)}&=& {1\over i} \left[\hat \phi[X](\sigma),\hat{\cal H}_\alpha[X](\sigma^\prime)\right],\\ 
\label{eq:Heq2}
{\delta \hat \pi[X](\sigma)\over\delta X^\alpha(\sigma^\prime)} &=& {1\over i} \left[\hat\pi[X](\sigma),\hat{\cal H}_\alpha[X](\sigma^\prime)\right],
\end{eqnarray}
where $\hat {\cal H}_\alpha[X]$ is the energy-momentum density of the scalar field at the slice embedded by $X$, normal-ordered in the operators $(\hat a_n^{(\pm)},  \hat a_n^{(\pm)\dagger})$. In null coordinates we have
\begin{equation}
\label{ }
\hat{\cal H}_\pm[X] = \pm {1\over 4} (X^{\pm\prime})^{-1}:(\hat \pi[X] \pm \hat \phi^\prime[X])^2 :.
\end{equation}
Just as the unitary transformation appearing in (\ref{eq:HU}) can be redefined by a (possibly embedding-dependent) phase factor, so one can also redefine the energy and momentum densities by the addition of a (possibly embedding-dependent) multiple of the identity without disturbing the validity of (\ref{eq:Heq1}), (\ref{eq:Heq2}).

In the (functional) \Schrodinger picture the canonical field operators are defined once and for all, independently of the embedding, and the state vectors are embedding-dependent. To construct this picture of dynamics we pick once and for all a fiducial embedding $X_0$ -- the initial slice -- upon which we identify the Heisenberg and \Schrodinger canonical field operators.  Given an initial state vector $|\Psi\>\in \F$ characterizing the state of the system at the time defined by $X_0$, the state vector $|\Psi,X\>\in \F$ on the embedding $X$ is defined by
\begin{equation}
\label{eq:SU}
|\Psi, X\> = \hat U[X_0,X] |\Psi\>.
\end{equation}
We remark that, as $\hat U$ is only determined up to a (possibly embedding-dependent) phase, the state vector on the slice defined by $X$ is determined only up to such a phase.

By considering deformations of the embedding in (\ref{eq:SU}), it follows  that the \Schrodinger picture state vector satisfies the following functional \Schrodinger equation \cite{TV1998}:
\begin{eqnarray}
\label{eq:FSeq}
i{\delta\over\delta X^\alpha(\sigma)} |\Psi,X\> &=& \left(\hat H_\alpha[X](\sigma) + A_\alpha[X](\sigma) \hat I\right)|\Psi,X\>\\
&\equiv& {\Ham}_\alpha[X](\sigma) |\Psi,X\>.
\end{eqnarray}
Here 
\begin{equation}
\label{ }
\hat H_\alpha[X] = \hat U[X_0,X] \hat {\cal H}_\alpha[X] \hat U^\dagger[X_0,X],
\end{equation}
represents the normal-ordered \Schrodinger picture energy-momentum densities associated with the slice embedded by $X$.  
We note that $\hat H_\alpha[X]$ is {\it not} normal-ordered in the $(\hat a,\hat a^\dagger)$ operators. Rather, $\hat H_\alpha[X]$ is normal-ordered in the \Schrodinger picture image of the creation and annihilation operators:
\begin{equation}
\hat b^{(\pm)}_n[X] = \hat U[X_0,X] \hat a_n^{(\pm)} \hat U^\dagger[X_0,X].
\end{equation}
In particular, this means the \Schrodinger image of the Fock vacuum state vector associated to the slice $X$, given by 
\begin{equation}
\label{ }
|\vac,X\>:=U[X_0,X]|\psi,0\>,
\end{equation} 
satisfies
\begin{equation}
\label{ }
\hat b^{(\pm)}_k[X]|\vac,X\> = 0,\quad k>0,
\end{equation} 
and
\begin{equation}
\label{ }
\<\vac, X|\hat H_\alpha[X]|\vac, X\> = 0.
\end{equation}
See Ref.~\onlinecite{TV1998} for an explicit expression for $|\vac,X\>$.

The $c$-number contribution to (\ref{eq:FSeq}), $A_\alpha=A_\alpha[X]$, is Kucha\v r's anomaly potential \cite{Kuchar1989}, and is explicitly computed from $\hat U$ in Ref.~\onlinecite{TV1998}.  Because $U[X_0,X]$ and hence $|\Psi,X\>$ is only determined up to a phase, the expression for $A_\alpha$ is likewise non-unique. We have
\begin{eqnarray}
\label{ }
|\Psi,X\> &\longrightarrow& e^{i\Lambda[X]}|\Psi,X\>,\nonumber\\ 
\Longrightarrow\quad\quad\quad&\quad&\nonumber\\
A_\alpha[X](\sigma) &\longrightarrow& A_\alpha[X](\sigma) -{\delta \Lambda[X]\over\delta X^\alpha(\sigma)}.
\end{eqnarray}

Given a foliation of $M$, {\it i.e.}, a 1-parameter family of embeddings $X(s)$, the functional \Schrodinger equation (\ref{eq:FSeq}) defines the \Schrodinger equation for evolution of the state vector $|\Psi(s)\> \equiv |\Psi, X(s)\>$ along the foliation. We have
\begin{equation}
\label{SE}
i {d\over ds} |\Psi(s)\> = \Ham[X(s)]|\Psi(s),
\end{equation}
where
\begin{equation}
\Ham[X(s)] = \int_0^{2\pi}d\sigma\, \dot X^\alpha(s) \Ham_\alpha[X(s)](\sigma).
\end{equation}

\section{\label{sec:section3} The bundle of state vectors}

Each normalized {\it state vector},
\begin{equation}
\label{ }
|\Psi\>\in\F,\quad \<\Psi|\Psi\> = 1,
\end{equation}
defines a pure quantum {\it state}, {\it i.e.,} a projection operator 
\begin{equation}
\label{ }
\hat P_\Psi = |\Psi\>\<\Psi|.
\end{equation}

Let us denote the set of pure quantum states by $\S$. Each element of $\S$ corresponds to a family of normalized vectors in $\F$, each element of the family differing by a phase.  The set $\Q$ of normalized state vectors in $\F$, equipped with the projection $\pi\colon \Q\to \S$ given by
\begin{equation}
\label{ }
\pi(|\Psi\>) = |\Psi\>\<\Psi|,
\end{equation}
defines a principal fiber bundle with structure group $U(1)$ \cite{BMNKZ}. The  action on $\Q$ of an element $e^{i\alpha}\in U(1)$ is 
\begin{equation}
\label{ }
|\Psi\> \to e^{i\alpha}|\Psi\>.
\end{equation}
A tangent vector to $\Q$ at $|\Psi\>$ is any vector $|t\>\in \F$ such that
\begin{equation}
\label{tangent}
Re(\<\Psi|t\>) = 0.
\end{equation}
A vertical tangent vector $|v\>$ to $\Q$ at $|\Psi\>$ satisfies $\pi_*(|v\>) = 0$ and is of the form
\begin{equation}
|v\> = i\alpha|\Psi\>,\quad \alpha\in {\bf R}.
\end{equation}

The bundle $\pi\colon \Q\to\S$ has a natural connection \cite{BMNKZ}. Horizontal tangent vectors $|h\>$ at $|\Psi\>$ are defined by this connection to satisfy
\begin{equation}
\label{ }
Im(\<\Psi|h\>)  = 0.
\end{equation}
In light of (\ref{tangent}), we see that horizontal vectors at $|\Psi\>$ are elements of $\F$ orthogonal to $|\Psi\>$.
The associated connection 1-form $\Omega$ defines a map from the tangent space to $\Q$ at $|\Psi\>$ to the Lie algebra $u(1)$ (the Lie algebra of imaginary numbers) given by 
\begin{equation}
|t\> \to \Omega(|t\>) = \<\Psi|t\>.
\end{equation}
We have, in particular,
\begin{equation}
\label{ }
\Omega(|v\>) = \Omega(i\alpha|\Psi\>) = i\alpha,
\end{equation}
and
\begin{equation}
\label{ }
\Omega(|h\>) = 0.
\end{equation}

\section{\label{sec:section5} The induced vacuum bundle}

The \Schrodinger image $|\vac,X\>$ on the slice embedded by $X$ of the Heisenberg vacuum state vector defines an embedding-dependent family of states
\begin{equation}
\label{ }
P_{vac}[X] = |\vac, X\>\<\vac, X|.
\end{equation}
We remark that while both the Heisenberg and \Schrodinger vacuum vectors are only determined up to a phase, the projection operator $P_{vac}[X]$ is uniquely determined.
$P_{vac}[X]$ defines a mapping from $\E$, the space of embeddings, into the space $\S$ of Fock states, 
\begin{equation}
P_{vac}\colon \E\to \S.
\end{equation}
There is then a canonically defined principal $U(1)$ bundle over $\E$ obtained by copying each fiber from $\Q$ over $P_{vac}[X]$ to sit over $X$. This is the induced (or ``pull-back'') bundle \cite{Steenrod}.  More precisely, the induced bundle $\B\to\E$ is defined as the subset of points $(X,|\Psi\>)\in\E\times \Q$ such that
\begin{equation}
\label{ }
P_{vac}[X] = |\Psi\>\<\Psi|.
\end{equation}
This means for each point $p=(X,|\Psi\>)$ in $\B$ there is a phase $e^{i\gamma_p}$ such that
\begin{equation}
\label{ }
|\Psi\> = e^{i\gamma_p}|\vac, X\>.
\end{equation}
The projection mapping $\Pi\colon \B\to\E$ is given by
\begin{equation}
\Pi(X,|\Psi\>) = X.
\end{equation}
The action of $e^{i\beta}\in U(1)$ on $\B$ is given by
 \begin{equation}
\label{ }
(X,|\Psi\>) \longrightarrow (X,e^{i\beta}|\Psi\>).
\end{equation}
Because the set $\E$ of orientation-preserving spacelike embeddings is contractible, the bundle $\B$ is trivial.  

As for any induced bundle, there is a bundle morphism $f\colon \B\to\Q$ satisfying
\begin{equation}
\label{fdef}
\pi\circ f = P_{vac}\circ \Pi.
\end{equation} 
In our case, this mapping is given by
\begin{equation}
\label{ }
f(X,|\Psi\>) = |\Psi\>.
\end{equation}
A tangent vector to $\B$ at $(X,|\Psi\>=e^{i\gamma}|\vac,X\>)$ is a pair $(V,|T\>)$, where $V\colon S^1\to T_XM$ is a deformation of an embedding,
\begin{equation}
\label{V}
V = \int_0^{2\pi} d\sigma\, V^\alpha[X](\sigma){\delta\over\delta X^\alpha(\sigma)},
\end{equation}
 and $|T\>$ is an element of $\F$ of the form
\begin{equation}
\label{eq:T}
|T\> = i\nu |\Psi\> +e^{i\gamma} |\Xi\>,
\end{equation}
where $\nu\in{\bf R}$ and
\begin{equation}
|\Xi\> = \int_0^{2\pi} d\sigma\, V^\alpha[X](\sigma) {\delta\over\delta X^\alpha(\sigma)}|\vac,X\>.
\end{equation}

\section{\label{sec:section6} The induced connection and the anomaly potentials}

Using the mapping $f\colon \B\to\Q$ defined in the previous section, we can pull back the connection 1-form $\Omega$ on $\Q$ to define an induced connection 1-form $\omega=f^*\Omega$ on $\B$. By definition, the action of $\omega$ on a tangent vector $(V,|T\>)$ at $(X,|\Psi\>)$ is given by the action of $\Omega$ on the push-forward of $(V,|T\>)$ at $f(X,|\Psi\>) = |\Psi\>$ on $\Q$. The push-forward is given by
\begin{equation}
\label{ }
f_*(V,|T\>) = |T\>.
\end{equation}
We then have (with $|T\>$ given by (\ref{eq:T}))
\begin{eqnarray}
\label{ }
\omega(V,|T\>) &=& \Omega(|T\>)\\
&=&\<\Psi|T\>\\
&=& i\nu + \<\vac,X|\Xi\>,
\end{eqnarray}
from which it follows the horizontal vectors $(V,|T\>)$ on $\B$ at $(X,|\Psi\>)$ have $|T\>$ orthogonal to $|\Psi\>$.

We are now ready to make explicit the link between the induced connection and the anomaly potential of Kucha\v r.  We do this by evaluating the 1-form $\omega$ in a gauge. A choice of gauge corresponds to a choice of cross-section $\rho\colon \E\to \B$, which can be defined by a mapping $\beta\colon \E\to{\bf R}$ via
\begin{equation}
\label{ }
 \rho(X) = (X,e^{i\beta[X]}|\vac, X\>).
\end{equation}
In light of the triviality of the induced bundle, this cross-section can be defined globally; for simplicity we assume this in what follows. We define the {\it gauge potential} to be the 1-form on $\E$ defined by $\A = i\rho^*\omega$.
Let $V$ be a tangent vector to $\E$ at $X$, as in (\ref{V}). The push-forward $\rho_*(V)$ is a tangent vector to $\B$ at $\rho(X)$  given by
\begin{equation}
\label{ }
\rho_*(V) = (V, \nabla_V(e^{i\beta[X]}|\vac, X\>)),
\end{equation}
where
\begin{equation}
\label{ }
\nabla_V(e^{i\beta[X]}|\vac, X\>) = \int_0^{2\pi} d\sigma\, e^{i\beta[X]}V^\alpha[X](\sigma)\left(  {\delta\over\delta X^\alpha(\sigma)}|\vac, X\>+ i{\delta\beta[X]\over\delta X^\alpha(\sigma)} |\vac, X\>\right).
\end{equation}
 At the point $X$ we have
\begin{equation}
\label{ }
\A(V) = i\omega(\rho_*V) = \int_0^{2\pi} d\sigma\, V^\alpha[X](\sigma)\left(i\<\vac, X|{\delta\over\delta X^\alpha(\sigma)}|\vac, X\>-{\delta\beta[X]\over\delta X^\alpha(\sigma)}  \right).
\end{equation}
Writing
\begin{equation}
\label{ }
\A(V) = \int_0^{2\pi} d\sigma\, \A_\alpha(\sigma) V^\alpha(\sigma),
\end{equation}
then using (\ref{eq:FSeq}) and taking account of  the normal-ordering of $H_\alpha$, we have
\begin{eqnarray}
\label{aa}
\A_\alpha(\sigma) &=& i\left\{\<\vac, X|{\delta\over\delta X^\alpha(\sigma)}|\vac, X\>\right\}-{\delta\beta[X]\over\delta X^\alpha(\sigma)}\\
&=&  A_\alpha(\sigma) -{\delta\beta[X]\over\delta X^\alpha(\sigma)}
\end{eqnarray}
Thus the anomaly potential coincides with the gauge potential of the connection $\omega$, {\it i.e.},  to the expression of the natural connection on the induced bundle in a particular gauge. 

Neither of the connections $\omega$ or $\Omega$ is flat. In particular, the gauge-invariant {\it field strength} $d\A$ -- which is $i$ times the pull-back to $\E$ of the curvature $d\Omega$ by $f\circ\rho$ -- is non-vanishing.  The components of $d\A$ correspond to the Schwinger terms in the commutator algebra of the operators $H_\alpha$.  This can be easily seen from our geometric construction as follows.  Let $U$ and $V$ be tangent vectors to $\E$:
\begin{equation}
\label{ }
U = \int_0^{2\pi}d\sigma\, U^\alpha[X](\sigma) {\delta\over\delta X^\alpha(\sigma)},\quad
V = \int_0^{2\pi}d\sigma\, V^\alpha[X](\sigma) {\delta\over\delta X^\alpha(\sigma)}.
\end{equation}
We have
\begin{eqnarray}
\label{ }
d\A(U,V)&=& i\int_0^{2\pi} d\sigma d\sigma^\prime \left[U^\alpha(\sigma)V^\beta(\sigma^\prime)-U^\beta(\sigma^\prime)V^\alpha(\sigma)\right]\nonumber\\ 
&\times&
\left\{\left({\delta\over\delta X^{\alpha}(\sigma)}\<\vac,X|\right)\left({\delta\over\delta X^{\beta}(\sigma^\prime)}|\vac, X\>\right)\right\} .
\end{eqnarray}
Using the functional \Schrodinger equation (\ref{eq:FSeq}), we then get
\begin{equation}
\label{schwing}
d\A(U,V) = i\int_0^{2\pi} d\sigma d\sigma^\prime U^\alpha(\sigma)V^\beta(\sigma^\prime)\Big(\<\vac,X|\Big[\hat H_\alpha(\sigma), \hat H_\beta(\sigma^\prime)\Big]|\vac,X\>\Big).
\end{equation}
The commutators of the normal-ordered operators $\hat H_\alpha$ close back into the $\hat H_\alpha$ up to embedding-dependent multiples of the identity --  the Schwinger terms \cite{Kuchar1989}:  
\begin{equation}
\left[\hat H_\alpha[X](\sigma), \hat H_\beta[X](\sigma^\prime)\right] = \int_0^{2\pi} d\sigma^{\prime\prime}\, C_{\alpha\beta}^\gamma[X](\sigma,\sigma^\prime, \sigma^{\prime\prime}) \hat H_\gamma[X](\sigma^{\prime\prime}) + {\cal F}_{\alpha\beta}[X](\sigma,\sigma^\prime)\hat I.
\end{equation}
See Ref.~\onlinecite{Kuchar1989} for explicit expressions.
 The Schwinger terms are then produced by the vacuum expectation value appearing in (\ref{schwing}) for $d\A$.

\section{\label{sec:section7}  Dynamical Phase, Geometric Phase and Foliation Independence of Time Evolution}


The integrability of (\ref{eq:FSeq}) implies that dynamical evolution between given initial and final slices is independent of the interpolating foliation \cite{Kuchar1989}. This can be seen directly using the geometric ingredients discussed in the previous sections. 

Consider a closed curve $\gamma$ in $\E$. Physically this corresponds to considering time evolution from an initial slice $X_0$ to a final slice $X_1$ along a foliation $X(s)$, $s\in [0,1]$ followed by evolution from $X_1$ to $X_0$ along a different foliation $\tilde X(s)$, $s\in [1, 2]$.  The mapping $P_{vac}$ (see (\ref{fdef})) defines a closed curve $\C=P_{vac}(\gamma)$ in $\S$, corresponding to the various \Schrodinger picture images of the Heisenberg vacuum state at each time slice labeled by $s$.  Because the quantum state only defines the state vector up to a phase,  there are infinitely many curves in $\Q$ associated to $\C$ -- these are the {\it lifts} of $\C$. These lifts need not be closed curves. We now consider a couple of lifts of $\C$ -- a {\it geometric lift} and a {\it dynamical lift} -- to demonstrate that (i) the geometric phase -- the holonomy of the connection $\omega$ -- corresponds to a foliation dependence in the evolution of state vectors generated by the \Schrodinger picture densities $\hat H_\alpha$ and (ii) the presence of the anomaly potential in (\ref{SE}) provides a dynamical phase which precisely cancels the geometric phase and removes the foliation dependence.  For simplicity we follow our earlier discussion and focus only on time evolution of the \Schrodinger picture vacuum state. It is straightforward to generalize the discussion to any Fock state.

The geometric lift \cite{BMNKZ} of $\C$, denoted by $|\Psi(s)\>_{geom}$, is the horizontal lift defined by the connection $\Omega$. This is not a closed curve; we have
\begin{equation}
\label{ }
|\Psi(2)\>_{geom} = e^{i\beta}|\Psi(0)\>_{geom},
\end{equation}
where the phase factor is the gauge-invariant holonomy of the connection $\Omega$ -- it is the geometric phase for this closed path in $\S$.  
It can be computed by integrating $\bf A$ ({\it i.e.}, the anomaly potential) around the closed curve $\gamma$ in $\E$, 
\begin{equation}
\label{ }
e^{i\beta} = \exp\left\{i \int_0^2 ds\, \dot \X^\alpha(s) {\bf A}_\alpha[\X(s)]\right\} = \exp\left\{i \int_0^2 ds\, \dot \X^\alpha(s)  A_\alpha[\X(s)]\right\}.
\end{equation}
where $\gamma$ is represented by the 1-parameter family of embeddings:
\begin{equation}
\label{ }
\X(s) =
\begin{cases}
{X(s),\ s\in[0,1]}\\
{\tilde X(s),\ s\in[1,2].}
\end{cases}
\end{equation}
That this phase is non-trivial follows from the non-triviality of the curvature of $\bf A$. As we have seen this curvature is determined by the Schwinger terms in the commutator algebra of $\hat H_\alpha$; see (\ref{schwing}).  Thus the Schwinger terms are responsible for the geometric phase.

The geometric lift obeys a \Schrodinger equation with a dynamical phase removed \cite{BMNKZ}:
\begin{equation}
\label{ }
i{d\over ds} |\Psi(s)\>_{geom} = \left(\hat \Ham[\X(s)] - E[\X(s)]\hat I\right)|\Psi(s)\>_{geom}.
\end{equation}
Here
\begin{equation}
\label{ }
\hat \Ham[\X(s)] = \int_0^{2\pi}d\sigma\, \dot \X^\alpha(s)\left\{\hat H_\alpha[\X(s)] + A_\alpha[\X(s)]\hat I\right\},
\end{equation}
and
\begin{equation}
\label{ }
E[\X(s)] = {}_{geom}\<\Psi(s)|\hat \Ham[\X(s)]|\Psi(s)\>_{geom} =  \int_0^{2\pi}d\sigma\, \dot \X^\alpha(s)A_\alpha[\X(s)],
\end{equation}
so that\begin{equation}
\label{ }
\hat \Ham[\X(s)] - E[\X(s)]\hat I =  \int_0^{2\pi}d\sigma\, \dot \X^\alpha(s)\hat H_\alpha[\X(s)] \equiv \hat H[\X(s)].
\end{equation}
Thus the geometric phase is the phase arising when considering cyclic evolution of the (vacuum) state vector generated by the normal-ordered \Schrodinger picture energy $\hat H[\X(s)]$ associated with the slicing.  The presence of the geometric phase implies a foliation dependence in the evolution of state vectors generated by  $\hat H$. The foliation dependence can be seen as follows. 

The unitary time evolution operator for the geometric lift going from $X_0$ to $X_1$ along the foliation $X(s)$ is given by
\begin{equation}
\label{ }
\hat U(0,1) =  {\cal T}\exp\left\{-i \int_0^1ds\, \hat H[X(s)]\right\}.
\end{equation} 
Here ${\cal T}\exp$ is the time-ordered exponential.   Similarly, the time evolution operator for the geometric lift going from $X_1$ to $X_0$ along the foliation $\tilde X(s)$ is given by
\begin{equation}
\label{ }
\hat U(1,2) = {\cal T}\exp\left\{-i \int_1^2ds\, \hat H[\tilde X(s)]\right\},
\end{equation}
We then have
\begin{equation}
\label{ }
e^{i\beta}|\Psi(0)\>_{geom} = \hat U(1,2) \hat U(0,1) |\Psi(0)\>_{geom},
\end{equation}
and it follows that
\begin{equation}
\label{ }
\hat U(0,1)|\Psi(0)\>_{geom} = e^{i\beta} \hat U^\dagger(1,2)|\Psi(0)\>_{geom}.
\end{equation}
Now, $\hat U^\dagger(1,2)$ is the time evolution operator for the geometric lift going  from  $X_0$ to  $X_1$ along the slicing determined by $\tilde X(s)$. We see then that the time evolution of the initial (vacuum) state vector generated by $\hat H$ along two different foliations connecting the same initial and final slices  leads to  final state vectors which differ by the geometric phase.

The dynamical lift is simply the curve $|\Psi(s)\>$ in $\Q$ solving the \Schrodinger equation defined from (\ref{eq:FSeq}) and the foliation $\X$ (see (\ref{SE})):
\begin{equation}
\label{ }
i{d\over ds}|\Psi(s)\> = \hat\Ham[\X(s)]|\Psi(s)\>.
\end{equation}
Of course, here we have
\begin{equation}
|\Psi(s)\> = |\vac,{\bf X}(s)\>.
\end{equation}
 Because of the integrability of (\ref{eq:FSeq}), the dynamical lift is a closed curve, {\it i.e.}, the net phase change in the dynamical lift vanishes: 
 \begin{equation}
|\Psi(2)\> = |\Psi(0)\>.
\end{equation}
This can be seen directly in terms of our previous discussion, thus showing the interplay between the anomaly potential and the connection on $\B$.  The time evolution generated by $\hat\Ham$   along $\X(s)$ is given by
\begin{equation}
|\Psi(2)\>  =   {\cal T}\exp\left\{-i\int_0^2 ds\, \dot {\bf X}^\alpha(s) \left(\hat H_\alpha[{\bf X}(s)] + A_\alpha[{\bf X}(s)] \hat I\right)\right\} |\Psi(0)\>.
\end{equation}
It is permissible to factor out the exponential  of the anomaly potential since the exponent is a multiple of the identity. We then get
\begin{eqnarray}
|\Psi(2)\>& = & \exp\left\{-i\int_0^2 ds\, \dot {\bf X}^\alpha(s) A_\alpha[{\bf X}(s)]\right\}{\cal T}\exp\left\{-i\int_0^2 ds\, \hat H[\X(s)] \right\}|\Psi(0)\>\\
 & = &   \exp\left\{-i\int_0^2 ds\, \dot {\bf X}^\alpha(s) A_\alpha[{\bf X}(s)]\right\} e^{i\beta} |\Psi(0)\>\\
 & = & |\Psi(0)\>.
\end{eqnarray}
Thus the anomaly potential leads to a dynamical phase which cancels the geometric phase caused by the Schwinger terms in the algebra of energy-momentum densities. Reasoning as before, we see that the dynamical evolution generated by $\hat\Ham$ is foliation independent.

\end{document}